# Learning to Gather without Communication

El Mahdi El Mhamdi  
EPFL  
elmahdi.elmhamdi@epfl.ch

Rachid Guerraoui  
EPFL  
rachid.guerraoui@epfl.ch

Alexandre Maurer  
EPFL  
alexandre.maurer@epfl.ch

Vladislav Tempez  
ENS Rennes  
vladislav.tempez@ens-rennes.fr

*Abstract*—A standard belief on emerging collective behavior is that it emerges from simple *individual rules*. Most of the mathematical research on such collective behavior starts from imperative individual rules, like *always go to the center*. But how could an (optimal) individual rule emerge during a short period within the group lifetime, especially if communication is not available.

We argue that such rules can actually emerge in a group in a short span of time via collective (multi-agent) reinforcement learning, i.e learning via rewards and punishments.

We consider the *gathering* problem: several agents (social animals, swarming robots...) must gather around a same position, which is not determined in advance. They must do so without communication on their planned decision, just by looking at the position of other agents.

We present the first experimental evidence that a gathering behavior can be *learned* without communication in a partially observable environment. The learned behavior has the same properties as a *self-stabilizing* distributed algorithm, as processes can gather from any initial state (and thus tolerate any transient failure). Besides, we show that it is possible to tolerate the brutal loss of up to 90% of agents without significant impact on the behavior.

## I. INTRODUCTION

Collective behavior such as shoaling and schooling in fishes, herding in mammals or flocking in birds, are striking examples of complex patterns emerging from networked distributed behaviors. Computer Science and more specifically Artificial Intelligence investigated the emerging behaviors in such *multi-agent* systems [36]. These systems are appealing for their robustness and adaptability [30]. However, their distributed nature makes it difficult to determine the actual algorithm controlling each agent [27], and thus to implement it and leverage it in various contexts.

A solution is to use *machine learning* [41] to capture the desired behavior. Machine learning allows a program to automatically extract a model from a data set or from its interactions with the environment: one does not need to explicitly specify the algorithm controlling the agents. *Reinforcement learning* [45], [42] is the specific machine learning paradigm that enables to obtain a desired behavior with the simplest feedback from the environment. It is particularly useful in network related problems [37], [7], [23]. In short, reinforcement learning consists, for the program, in receiving *rewards* and *penalties* from the environment, and learning which behavior leads to rewards and which behavior leads to penalties.

Here, we study the fundamental problem of *gathering* [13], [4], [15], [3]: several agents must gather around a same position which is not determined in advance. Agents can "see" each other, but are not allowed to directly communicate with signals or messages. So far, this problem has been solved with explicit algorithms (i.e., a human explicitly writes an algorithm, and one shows that it works). However, the question whether the agents can *learn* to gather with only simple rewards and penalties from the environment (and without communication) remained open.

In this paper, we present the first experimental evidence that the answer to this question is affirmative: agents can indeed learn a gathering behavior. The system controlling these agents is fully decentralized and embodied in each agent. Communication is not allowed and agents can only obtain the position of other agents relatively to their own. The agents are rewarded for being in a group and penalized for being isolated. The learned behavior has the same properties as a self-stabilizing distributed algorithm [16], as processes can gather from any initial state.

We show that agents can learn to gather on a one-dimensional ring. This result can be immediately generalized to higher dimensions, if the gathering occurs independently in each dimension. For instance, the product of two one-dimensional rings is a two-dimensional torus, where the agents gather at a same $x$ position – and, in parallel and independently, at a same $y$ position.

A technical difficulty lies in the "combinatorial explosion" of the number of states (induced by the fact that there is not one, but *many* agents). To overcome this difficulty, the agents approximate the environment by grouping close positions into clusters: each agent only perceives an *approximation* of the distribution of other agents in each cluster. This enables to keep the learning space constant (i.e., independent of the number of agents and the size of the ring). We show that, surprisingly, the agents manage to gather almost perfectly despite this rough approximation.

We then consider the problem of increasing the number of agents. A natural belief would be that the agents have to *re-learn* to gather in this case. Interestingly, we show that the learned behavior can directly apply to a much larger number of agents – namely, if agents have learned to gather in groups of 10, we show that they immediately know how to gather in groups of up to 100. Aside from saving learning time, the interest of this approach is that such a group of 100 agents is inherently and deeply *robust* (fault-tolerant), because it can tolerate the loss of up to 90 agents. We also compare the learned behavior with a deterministic algorithm that moves towards the barycenter of the agents. We thus show that, even with a relatively simple learning scheme, we can reach the same performances as this deterministic behavior.

Our source code for all simulations, data generation and display is available in [1], together with the implementation of the learning process.

The paper is organized as follows. In Section II, we give a state of the art of reinforcement learning in multi-agent systems w.r.t. the gathering problem. In Section III, we present the Q-learning technique (with eligibility trace), then a precise formulation of the gathering problem in a Q-learning framework. In particular, we describe which state and actions are used to model the gathering problem in Q-learning. In Section IV, we explicit the numerical parameters used to implement our model. For pedagogical reasons, we first present results for a default setting; then, we show that the learned behaviors can be reused with more agents. In Section V, we summarize our results and discuss various possible extensions of this work.

## II. STATE OF THE ART

Reinforcement learning [42], [22] consists in taking simple feedback from the environment to guide learning. The general idea is to associate rewards and penalties to past situations in order to learn how to act in future ones. The principle differs from that of supervised learning [18], [22] by the nature of the feedback. In supervised learning, an agent is taught how to perform precisely on several examples. In reinforcement learning, the agent only gets an appreciation feedback from the environment. For instance, in dog training, dogs are rewarded when doing correct actions and punished when behaving badly. The advantage here is the possibility to have a feedback in situations where the correct behavior is unknown. Several successful AI approaches use reinforcement learning, one spectacular example being the performance of AlphaGo [40] defeating the world Go champion Lee Sedol.

So far, reinforcement learning has mainly been used in situations with only one learning agent (*single-agent* systems), with important results [20], [17], [19], [24], [31], [38].

*Multi-agent* systems involve numerous independent agents interacting with each other. Many works on multi-agent reinforcement learning consider problems where only 2 or 3 agents are involved [6], [8], [14], [29], [35], [44], [47]. Some deal with competitive games (e.g. zero-sum games) [2], where agents are rewarded at the expense of others. Other tackle collaborative problems, but the reward is global and centralized [43].

In general, communication mechanisms are used to share information among agents [9], [25], [26], [33], [37], [39], [48] in order to increase the learning speed. Still, in some cases, communication between independent agents is difficult, impossible, or at least very costly [46], [5]. In these situations, it might be useful to devise a learning process that does not rely on communication.

Yet, so far, very few approaches considered a genuinely distributed setting where each agent is rewarded individually, and where agents do *not* communicate. In [32], the problem and the constraints are similar to our work, but the rewards are given for taking an *action* instead of reaching a *state*. Consequently, the final behavior is predetermined by the model itself. In [10], even if the constraints are similar (cooperative task, no communication and individual rewards), the problem tackled is fundamentally different: the task only requires the cooperation of agents by groups of two (not of all agents simultaneously).

## III. MODEL

### A. Q-learning

As recalled in the previous section, the goal of reinforcement learning is to make agents *learn* a behavior from reward-based feedback. In this paper, we work with a widely used reinforcement learning technique called *Q-learning* [45], [42], [47], [25], [11], [17]. More specifically, we use Q-learning with *eligibility trace* [28], [42] as explained in what follows.

Q-learning was initially devised for single agent problems. Here, we consider a multi-agent system where each agent has it own learning process. We describe in the following the learning model of *one* agent taken independently.

Let $A$ be a set of *actions*, and let $S$ be a set of *states* (representing all the situations in which the agent can be). The sets $A$ and $S$ contain a finite number of elements. In each state $s$, the agent may chose between different actions $a \in A$. Each action $a$ leads to a state $s'$, in which the agent receives either a positive reward, a negative reward or no reward at all. The objective of Q-learning is to compute the cumulative expected reward for visiting a given state. Intuitively, this is materialized by the fact that *learning*, in Q-learning, is all about updating the Q-value using the mismatch between the previous Q-value and the observed reward.

Let $\pi : S \to A$ be the *policy function* of an agent – i.e., a function returning an action to take in each state.

Let $X_t^{\pi,s_0}$ be the state in which the agent is after $t$ steps, starting from state $s_0$ and following the policy $\pi$. In particular, $X_0^{\pi,s} = s$.

Let $r : S \to \mathbb{R}$ be the *reward function* associating a reward to each state.

The *cumulative expected reward* over a period $I = [\![0, N]\!]$ of state $s$ is

$$\sum_{t \in I} \mathbb{E}(\gamma^t r(X_t^{\pi,s}))$$

where $\gamma \in [0, 1]$ is a *discount parameter* modulating the importance of long term rewards. The long term rewards become more and more important when $\gamma$ is close from $1$.

When predicting the best transition from one state to another (by taking a given action) is difficult or impossible, it is useful to compute a cumulative expected reward of a couple $(s, a)$.

Under the assumption that each couple $(s, a)$ is visited an infinite number of time, it is possible, following the law of large numbers, to estimate without bias the expected cumulative reward by sampling [45], i.e by trying state-action couples and building an estimator of the expected reward. We denote this estimator $Q(s,a)$, and call it the *Q-value* of the state-action couple $(s,a)$. The following formula is the usual update rule to compute an estimator of the Q-value.

$$Q_{t+1}(s,a) = (1-\eta)Q_t(s,a) + \eta(r(X_{t+1}) + \gamma \max_{a'}(Q_t(s,a')))$$

If action $a$ is taken in state $s$ at step $t$.

$$Q_{t+1}(s',a') = Q_t(s',a')$$

otherwise.

Here, $\eta$ is a parameter called the *learning rate* that modulates the importance of new rewards over old knowledge. $Q_t$ is the estimate of the cumulative expected reward after $t$ samples.

A complementary approach to get better estimations of Q-values with fewer samples is to use *eligibility trace* [28], [42]. The idea is to keep trace of older couples $(s,a)$ until a reward is given, and to propagate a discounted reward to the couples $(s,a)$ that led to the reward several steps later. Formally, for each state $s$, a value (eligibility) $e_t(s)$ is attributed. $e$ is initialized at $e_0(s) = 0$ for every state $s$ then updated as follows:

$$e_{t+1}(s) = \gamma\lambda * e_t(s) + 1$$

if $s_t = s$.

$$e_{t+1}(s) = \gamma\lambda * e_t(s)$$

otherwise. Using the eligibility trace, Q-values are updated by the scaling the update rule described above with eligibility values. The factor $\gamma\lambda$ used in the update of the eligibility acts as a discount in time: older visited states get less reward than recent visited states.

In addition to update rules for learning, we need a policy for choosing actions. An $\epsilon$-greedy policy $\pi$ is a stochastic policy such that: (1) with probability $(1-\epsilon)$, $\pi(s) = a$ when $(s,a)$ yields the highest expected cumulative reward from state $s$, and (2) with probability $\epsilon$, a random action is chosen in $A$. The parameter $\epsilon$ is called the *explore rate* and modulates the trade-off between *exploration* of new and unknown states (to obtain new information) and *exploitation* of current information (to sample valuable states more precisely and thus be rewarded).

*B. Setting*

We consider a *ring* topology. This is a simple topology for a bounded space that avoids non-realistic borders effects (i.e no need to "manually" replace an agent in the middle of the states-space if the agent reaches the border in the case of a square for example). There are $n$ positions $\{0, \ldots, n-1\}$. $\forall k \in \{0, \ldots, n-2\}$, positions $k$ and $k+1$ are adjacent, and positions $n-1$ and $0$ are also adjacent. Each agent has a given position on the ring. This space has only one dimension, but our results may be extended to higher dimension spaces by applying the approach independently on each dimension.

The time is divided into discrete steps $1, 2, 3, \ldots$. At the beginning of a given step $t$, an agent is at a given position. The possible actions are: *go left* (i.e. increase position), *go right* (i.e. decrease position) or *do not move*.

The current state of each agent is determined by the relative positions of other agents. However, we cannot associate a state to each combination of position of other agents, because of "combinatorial explosion". Thus, in order to limit the maximal number of states, each agents perceives an *approximation* of the positions of other agents. Besides, a state must not depend on the number of agents, in order to have a scalable model and to tolerate the loss of agents.

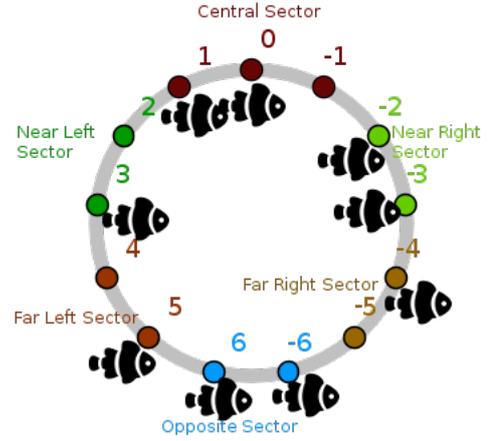

Fig. 1. Default sector delimitation (on a ring of size 13). The *Central* sector contains 3 positions centered around 0 (i.e., the position of the current agent). *Near* sectors contain two positions each, adjacent to the Central sector. Same for *Far* sectors and the *Opposite* sector.

Thus, our state model is the following. The space is divided into groups of close positions called *sectors*. Each agent does not perceive the exact number of agents per sector, but the *fraction* of the total population in each sector. A *state* is given by the knowledge of the fractions of the total population in each sector with a precision of 10% (i.e. the possible values are multiples of 10%, rounded so that the sum of the fractions equals 100%). The choice of 10% precision here is an arbitrary value to reduce computational cost, this value can be optimized as an hyper-parameter.

The delimitation of the sectors is not absolute but *relative* to the position of each agent: each agent has its own sector delimitation centered around itself.

This delimitation is set to 6 sectors (as for the precision value of 10% described above, this choice can be left as an hyper-parameter, but optimizing it is out of the scope of this work). The first sector is centered around the agent position (its size corresponds to the size of the neighborhood where we expect the other agents to gather). This sectors is the *Central* sector. The agents in the central sector of a given agent are called its *neighbors*. Two more sectors are adjacent to the central sector, the *Near Right* and *Near Left* sectors. The *Far left* and *Far Right* are a second layer after the near sectors. Finally, the *Opposite* sector is the sector diametrically opposed to the Central one. The exact size of each sector is a parameter of the problem, as well as the number of agents and the number of positions.

An example of sectors delimitation is given in Figure 1, for a ring of size 13.

## C. Rewards

Each agent is rewarded if it has a large enough number of neighbors (i.e., more than a certain fraction of the total population is in its central sector). Each agent is penalized if it has not enough neighbors (i.e., less than a certain fraction of the population is in its central sector).

## D. Learning process

The learning phase is organized as follows:
- The initial positions of agents are random, following a uniform distribution.
- At each step, each agent decides where to go with a $\epsilon$-greedy policy.
- When all the decisions are taken, all the agents move simultaneously.
- After moving, they consider their environment, get rewards and update their Q-values with respect to these rewards.
- The learning phase is subdivided in *cycles* of several steps. At the end of each cycle, the position of agents is reset to random positions. This ensure that the environment is diverse enough to learn a robust behavior. After position reset, the agents can move again for another cycle.

The duration of a cycle is set proportional to the size of the ring (e.g. 5 times the size of the ring) in order to give enough time to the agents to gather: this time depends on the distance they have to travel, and this distance depends on the size of the ring. To update Q-values, Q-learning with eligibility traces is used. Eligibility traces are reset at the end of each cycle, and each time, a reward is given to an agent.

## E. Problem

Intuitively, the goal is to make the agents learn a *gathering* behavior, that is: within a reasonable time in a same cycle, the agents become (and remain) reasonably close to each other. This criteria is voluntarily informal, and its satisfaction will be evaluated with several metrics in the next section.

More precisely, the problem consists in computing, for each agent, a value $Q(s, a)$ for each couple state-action $(s, a)$. This value indicates which action $a$ to take in state $s$ in order to increase the likelihood of obtaining a reward. Our objective is to verify experimentally that the $Q$-values learned in this fashion lead to an efficient gathering of the agents.

## IV. RESULTS

We consider a ring of size 13, with a sector division such as described in Figure 1. A *group* exists if at least one agent has more than 80% of the population as neighbors. An agent is given a reward of value 100 if the fraction of neighbors is more than 80% of the population, and a penalty of value $-5$ if it is 10% or less.

The exploration rate is $\epsilon = 0.1$, the learning rate is $\eta = 0.1$ and the discount factor is $\gamma = 0.95$. The duration of a cycle is 65 steps (around 5 times the size of the ring), and the duration of the learning phase is 5000 cycles.

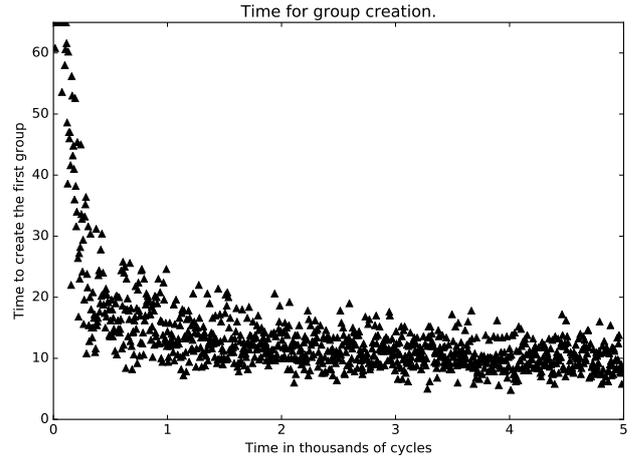

Fig. 2. Time needed to form a group from random initial positions for 10 agents on a ring of size 13. Each point is an average over 5 cycles.

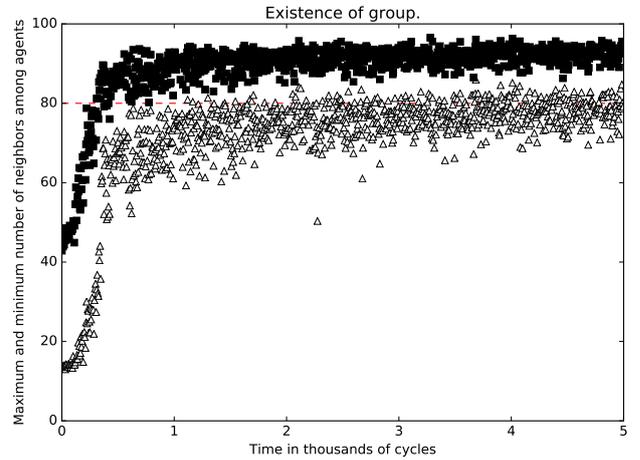

Fig. 3. Maximum and minimum number of neighbors, in percent of the total population, during learning, for 10 agents on a ring of size 13. Maximum is black squares and minimum is white triangles. The dashed line is the minimum number of neighbors needed to be considered in the group: 80% of the total number of agents. Each point is an average over 325 steps, including time before creation of the first group.

## A. Results for 10 agents

We first consider a population of 10 agents. To assess the quality of the learned behavior, we compute several metrics. We first consider the time needed to form a group from random initial positions, and see how it evolves during the learning phase. Then, to ensure that groups are not only formed but also maintained, we observe the evolution of the number of neighbors among the population. To evaluate the learning qualitatively, we look at the exact behavior of agents at the beginning, middle and end of the learning phase. Finally, we study the impact of a longer learning phase.

**Time to form a group.** Figure 2 shows the time that agents need to gather and form the first group (i.e., at least one agent is rewarded), starting from random initial positions. We observe that this time decreases during the learning phase and stabilizes around 10 steps.

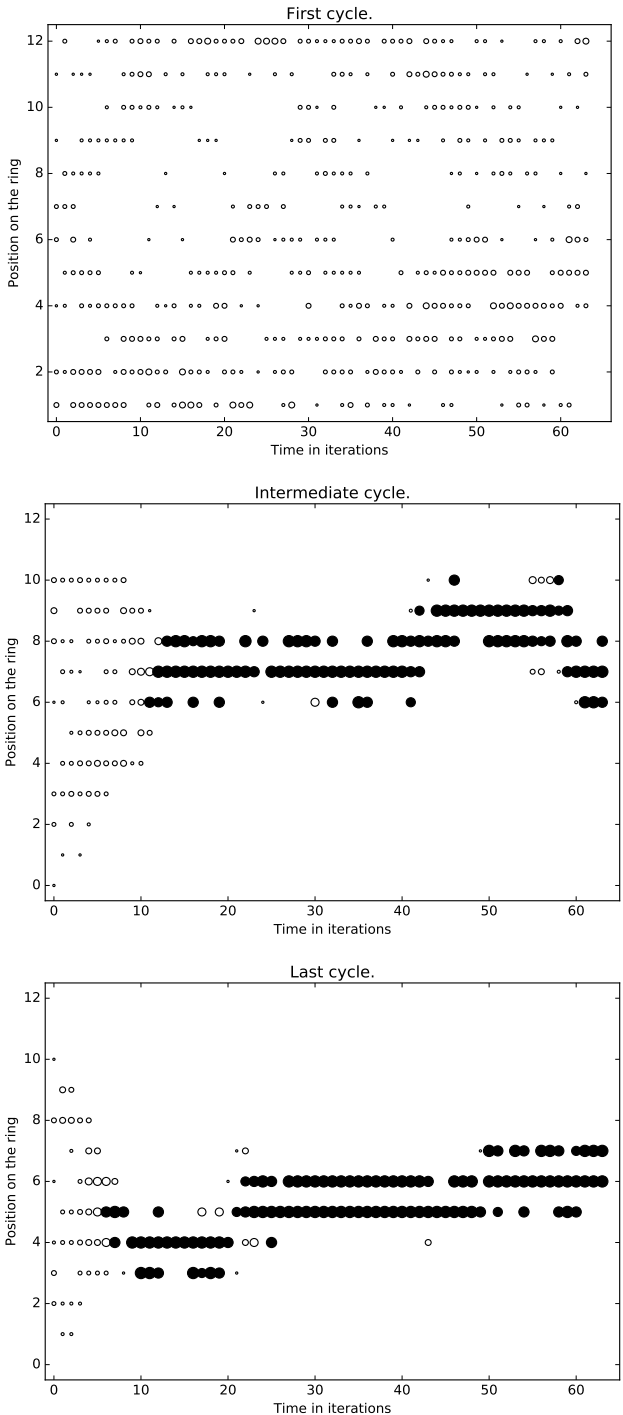

Fig. 4. Evolution over time of the number of neighbors at each position of the ring during a cycle. Larger dots represent a higher number of neighbors. Positions where agents are considered to be in the group are in black, others in white.

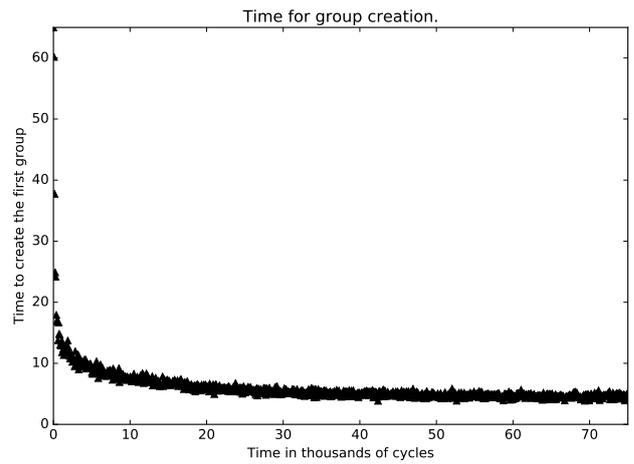

Fig. 5. Time needed to form a group from random initial position for 10 agents on a ring of size 13. Each point is an average over 75 cycles. Learning phase is 75 000 cycles long.

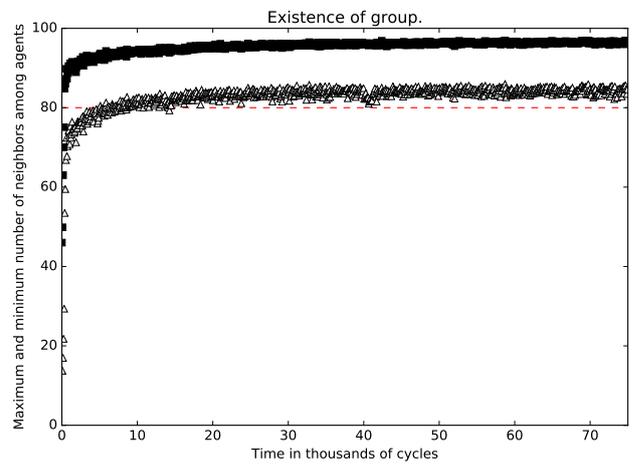

Fig. 6. Maximum and minimum number of neighbors, during learning, for 10 agents on a ring of size 13. Maximum is black squares and minimum is white triangles. The dashed line is the minimum number of neighbors needed to be considered in the group (i.e. 80% of the total number of agents). Each point is an average over 75 cycles (4875 steps), including time before creation of the first group. The learning phase is 75 000 cycles long.

**Number of neighbors.** Figure 3 shows the minimal and maximal number of neighbors over all agents. When the maximal number of neighbors is above 80%, it means that a group exists. When the minimal number of neighbors is above 10%, it means that no agent is isolated; when it is above 80%, it means that all agents are in the group. We observe that the agents learn, not only to gather, but also to *maintain* the group and avoid being isolated. Indeed, the maximum number of neighbors is higher than 80% of the total number of agents, and the minimum is higher than 10%. We also observe that the minimum number of neighbors is close to 80% at the end of the learning phase. It means that even the agents that are not *always* in the group are often in it.

Note that these average values include the iterations starting from the beginning of each cycle, where the agents are not yet gathered (i.e. around 10 iterations at the end of the learning phase).

**Qualitative evolution.** Figure 4 contains three plots that show the qualitative evolution of the learning for three cycles, at the beginning, middle and end of the learning phase.

In the first figure (beginning of the learning phase), we observe that the agents are quite uniformly distributed: the circles are white and small, indicating few neighbors and no significant group formation.

In the second figure (middle of the learning phase), we observe that the agents converge to a same position, forming a group in approximately 10 steps. The large black circle indicate that at least 80% of the total number of agents are neighbors of the position, i.e. that a group exists. We can see that this group is maintained after its formation until the end of the cycle. We also observe that the group itself is slowly moving during the cycle, while being maintained. We notice that there are very few agents outside the group after its formation.

In the third figure (end of the learning phase), we observe that agents still converge to form a group, but the group is formed earlier than before (around 7 steps). The group is still maintained and still moves during the cycle. We can notice even less agents outside the group than before.

**Longer learning phase.** We finally study the impact of a longer learning phase: 75 000 cycles instead of 5000.

Figure 5 is the equivalent of Figure 2 for a longer learning phase. At the end of the learning, the agents are gathering faster (around 5 steps) and are less often outside of the group.

Figure 6 is the equivalent of Figure 3 for a longer learning phase. We observe that the minimum number of neighbors goes above 80%, which means that all the agents are in the group most of the time.

*B. Scalability and comparison with a deterministic algorithm*

In the section, we explore the scalability and robustness properties of the aforementioned learning scheme. We show that the agents that have learned Q-values with default parameters in 75 000 cycles are able to gather with more agents *without* any new learning: we can take several agents that have learned in groups of 10 until we obtain a group of 100.

In a second time, we compare this behavior with a *deterministic* gathering algorithm (i.e., where the behavior in known in advance and not learned).

- First, we compare the learned behavior to an algorithm that uses the *exact* and *absolute* positions of all the agents (by opposition to relative positions and approximations used during learning). With this algorithm, agents always move towards the barycenter [12], [34] of all the agents. As this algorithm has an exact view on the environment, the performances are better (but not way better).

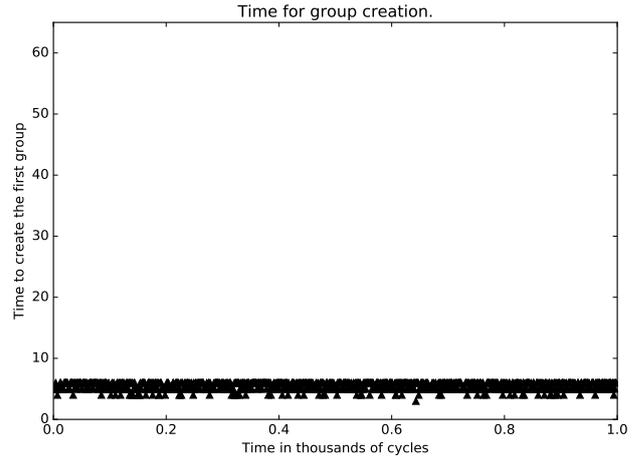

Fig. 7. Time needed to form a group from random initial positions for 100 agents on a ring of size 13 with (deterministic algorithm). Average is 5.4 steps, median is 5.0 steps and standard deviation is 0.6.

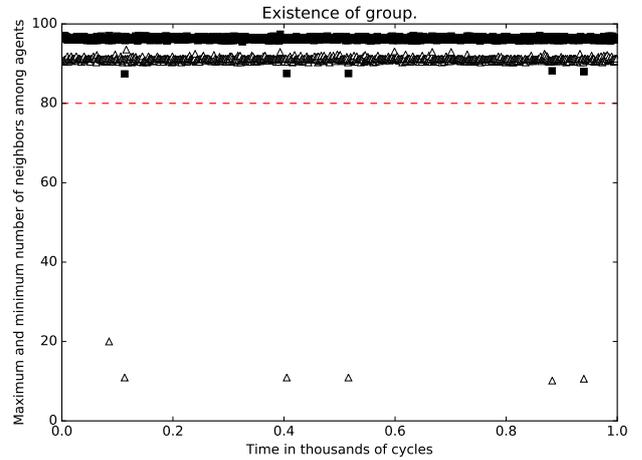

Fig. 8. Maximum and minimum number of neighbors for 100 agents on a ring of size 13 (deterministic algorithm). Maximum is black squares and minimum is white triangles. The dashed line is the minimum number of neighbors needed to be considered in the group. Each point is an average over a cycle (65 steps). Average is 90.6%, median is 91.1% and standard deviation is 6.1% for the minimum number of neighbors. Average is 96.3%, median is 96.3% and standard deviation is 0.7% for the maximum number of neighbors.

- We then make a fairer and more meaningful comparison with an algorithm that uses the same perceptions as the learning algorithm. With an equally constrained perception of the environment, we get results that are similar to the learned algorithm (the learned algorithm even slightly better in terms of "time to form a group"). We thus show that, even with a relatively simple learning scheme, we can reach the same performances as a deterministic behavior.

Note that, since the agents have already learned a behavior, there is no more "progression" visible on the plots.

**Time to create a group for 100 agents.** On Figure 9, we can see the time needed to form a group for 100 agents on a ring of size 13. Compared to the case with 10 agents, the time needed to form a group including 80% of the population is higher (around 10 steps in average). But the agents are still able to gather in a short time (the worst case is no more than 50 steps) most of the time: 997 times over 1000.

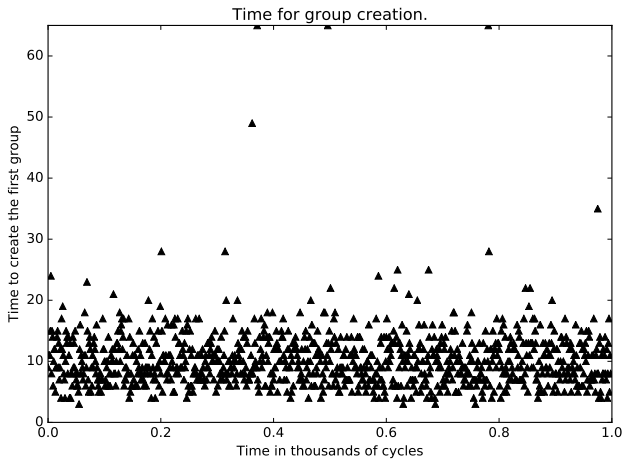

Fig. 9. Time needed to form a group from random initial positions for 100 agents on a ring of size 13 (learned behavior). Average is 10.4 steps, median is 10.0 steps and standard deviation is 5.1.

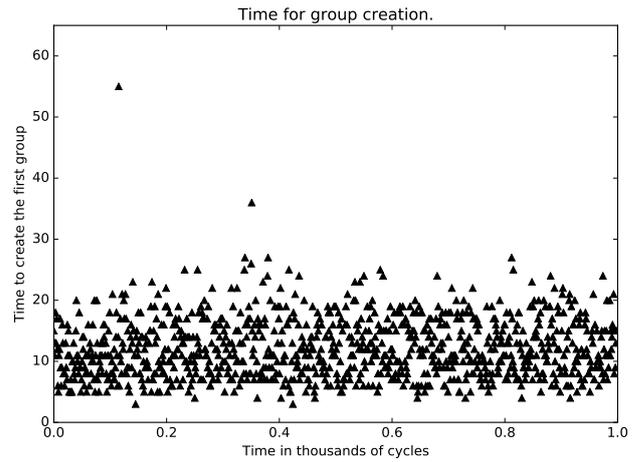

Fig. 10. Time needed to form a group from random initial positions for 100 agents on a ring of size 13 (Q-deterministic algorithm). Average is 12.1 steps, median is 11.0 steps and standard deviation is 4.9.

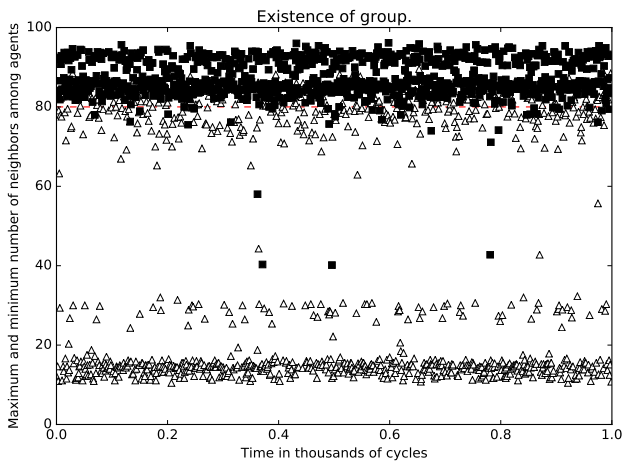

Fig. 11. Maximum and minimum number of neighbors for 100 agents on a ring of size 13. Maximum is black squares and minimum is white triangles (learned behavior). The dashed line is the minimum number of neighbors needed to be considered in the group. Each point is an average over a cycle (65 steps). Average is 40.4%, median is 16.3% and standard deviation is 31.0% for min neighbor. Average is 87.1%, median is 86.0% and standard deviation is 5.1% for max neighbor.

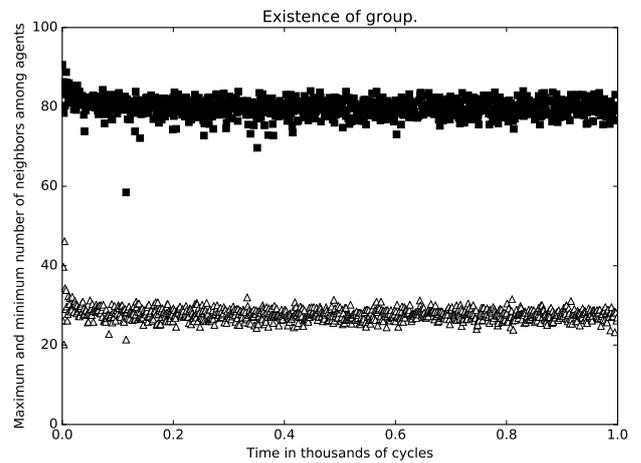

Fig. 12. Maximum and minimum number of neighbors for 100 agents on a ring of size 13 (Q-deterministic algorithm). Maximum is black squares and minimum is white triangles. The dashed line is the minimum number of neighbors needed to be considered in the group. Each point is an average over a cycle (65 steps). Average is 27.8%, median is 27.8% and standard deviation is 2.3% for the minimum number of neighbor. Average is 79.9%, median is 80.0% and standard deviation is 2.3% for the maximum number of neighbor.

**Number of neighbors for 100 agents.** On Figure 11, we observe that the maximum number of neighbors is higher than 80% most of the time, which means that a group exists most of the time. We also observe that the minimum number of neighbors is often low. This means that a few agents, even if not isolated, are unable to join the main group.

**Performances of the deterministic algorithm.** On Figure 7[1] and 8, we can observe that the deterministic algorithm is better than the learned behavior. In average, the agents gather in 5 steps with a standard deviation of 0.6. Moreover, the maximum and minimum number of neighbors are very high (average: resp. 96% and 91%). However, these good results are only possible because this algorithm uses the exact and absolute positions of other agents.

---
[1]Note that the figures are intentionally numbered to keep figures 9 and 10 (resp. 11 and 12) side by side, in order to have a clearer comparison between these figures.

**Fairer comparison.** To make a fairer comparison between deterministic algorithm an learned behavior, we try to impose to the deterministic algorithm the same constraints that were imposed to the learning algorithm: relative position, sector approximation and action choice with Q-values. To do so, we compute Q-values with the help of the deterministic algorithm. Each agent decide how to act according to the deterministic algorithm, and Q-values are computed along the sequence of actions determined by the deterministic algorithm. It allows each agent to compute Q-values for couples $(s, a)$ of states and actions. We call this algorithm the *Q-deterministic algorithm*: the desired behavior is known in advance, but we imposes the same perception constraints to the agents than the learned behavior.

In Figure 10, we observe that the time needed to form a group has the same distribution as the learned behavior in Figure 9. The average time is even slightly better for the learned behavior (10 steps) than for the Q-deterministic algorithm (12 steps). However, the standard deviation is slightly higher for learned behavior (5.1) than for the Q-deterministic algorithm (4.9).

In Figure 10, we represent the distribution of the number of neighbors. Here again, we observe that the distribution is better for the learned behavior (Figure 11) than for the Q-deterministic algorithm (Figure 12): the average of the maximum number neighbors is better (87% versus 80%) as well as the average of the minimum number of neighbors (40% versus 28%)[2]. However, the distribution of the number of neighbors is more sparse for the learned behavior.

## V. CONCLUSION

In this paper, we showed for the first time that it was possible for independent agents to learn a gathering behavior without communications via Q-learning. We implemented Q-learning on independent and non-communicating agents evolving on a one-dimensional ring topology. We designed approximations for environment and other agents perceptions in order to reduce the size of the learning space and make the behavior independent of the number of agents. We obtained an efficient and fast gathering behavior from random initial positions for 10 agents on a ring of size 13. After learning, the agents gather quickly (around 5 steps in average) and then maintain the cohesion of the group. Moreover, this learned behavior is scalable: it was possible to use it on 100 agents gathering on a ring of size 13 with no further learning. The quality of the gathering behavior for 100 agents is not as good as for 10 agents, but the average time to gather remains good (around 10 steps in average) and the worst time is less than 50 steps. Moreover, the learned behavior is a bit better than a deterministic algorithm equally constrained. This scalability, combined to the perception approximations, also allows robustness: the agent are capable of gathering independently of the number of other agents, if this number is between 10 and 100. The learned behavior is robust in the sense that a system starting with 100 agents can tolerate the loss of up to 90 agents. Overall, this work shows that Q-learning can be used in multi-agent, non fully observable environments without communication with other agents.

---

[2]Many light triangles are between 60% and 80% on Figure 11, which explains the higher average value.

In order to extend this work, it might be interesting to investigate how this multi-agent behavior emerges from the individual behavior of each agent, the difference of behavior between agents, and to quantify the importance of diversity in the behavior of agents.

Another track to continue this work would be to devise a way for agents to design or learn their *own* approximations of their environment. This could be done through unsupervised learning [21], or with the help of the reward feedback from the environment (or by a combination of both). This automatic design of the perception approximation could allow to systematically find a good compromise between the reduction of the learning space and the capacity to perceive meaningful differences and learn complex tasks. Neural networks may be a good modular framework to model these approximations functions.

A major challenge would be to find a way to reuse the behavior learned with the old approximation, instead of relearning the behavior from scratch whenever a change occurs in the approximation. The relative dynamics of the two timescales (one for the evolution of the approximation, and one for the evolution of the behavior) would also be of a particular importance.

## REFERENCES

[1] Source code for the simulation of this paper. https://github.com/LearningToGatherWithoutCommunication/Ring, 2016.
[2] O. Abul, F. Polat, and R. Alhajj. Multiagent reinforcement learning using function approximation. *IEEE Trans. Syst., Man, Cybern. C*, 30(4):485–497, 2000.
[3] N. Agmon and D. Peleg. Fault-tolerant gathering algorithms for autonomous mobile robots. *SIAM Journal on Computing*, 36(1):56–82, 2006.
[4] H. Ando, Y. Oasa, I. Suzuki, and M. Yamashita. Distributed memoryless point convergence algorithm for mobile robots with limited visibility. *IEEE Transactions on Robotics and Automation*, 15(5):818–828, 1999.
[5] R. C. Arkin. Cooperation without communication: Multiagent schema-based robot navigation. *Journal of Robotic Systems*, 9(3):351–364, 1992.
[6] M. D. Awheda and H. M. Schwartz. Exponential moving average based multiagent reinforcement learning algorithms. *Artificial Intelligence Review*, 45(3):299–332, oct 2015.
[7] P. Beyens, M. Peeters, K. Steenhaut, and A. Nowe. Routing with compression in wireless sensor networks: a q-learning appoach. In *Proceedings of the 5th European Workshop on Adaptive Agents and Multi-Agent Systems (AAMAS)*, 2005.
[8] B. Bouzy and M. Métivier. Multi-agent learning experiments on repeated matrix games. In *Proceedings of the 27 th International Conference on Machine Learning,*, 2010.
[9] M. Brambilla, E. Ferrante, M. Birattari, and M. Dorigo. Swarm robotics: a review from the swarm engineering perspective. *Swarm Intell*, 7(1):1–41, jan 2013.
[10] O. Buffet, A. Dutech, and F. Charpillet. Shaping multi-agent systems with gradient reinforcement learning. *Auton Agent Multi-Agent Syst*, 15(2):197–220, jan 2007.
[11] L. Busoniu, R. Babuska, and B. D. Schutter. A comprehensive survey of multiagent reinforcement learning. *IEEE Transactions on Systems, Man, and Cybernetics, Part C (Applications and Reviews)*, 38(2):156–172, mar 2008.

[12] B. Charlier. Necessary and sufficient condition for the existence of a fréchet mean on the circle. *ESAIM: Probability and Statistics*, 17:635–649, 2013.

[13] M. Cieliebak, P. Flocchini, G. Prencipe, and N. Santoro. Solving the robots gathering problem. In *International Colloquium on Automata, Languages, and Programming*, pages 1181–1196. Springer, 2003.

[14] C. Claus and C. Boutilier. The dynamics of reinforcement learning in cooperative multiagent systems. *AAAI/IAAI*, (s 746):752, 1998.

[15] J. Czyzowicz, L. Gasieniec, and A. Pelc. Gathering few fat mobile robots in the plane. *Theoretical Computer Science*, 410(6):481–499, 2009.

[16] S. Dolev. *Self-Stabilization*. MIT Press, 2000.

[17] D. J. Finton. When do differences matter? on-line feature extraction through cognitive economy. *Cognitive Systems Research*, 6(4):263–281, dec 2005.

[18] J. Friedman, T. Hastie, and R. Tibshirani. *The elements of statistical learning*, volume 1. Springer series in statistics Springer, Berlin, 2001.

[19] M. R. Glickman and K. P. Sycara. Evolutionary search, stochastic policies with memory, and reinforcement learning with hidden state. In *ICML*, 2001.

[20] X. Guo, S. Singh, H. Lee, R. Lewis, and X. Wang. Deep learning for real-time atari game play using offline monte-carlo tree search planning. In *Proceedings of the 27th International Conference on Neural Information Processing Systems*, NIPS'14, pages 3338–3346, Cambridge, MA, USA, 2014. MIT Press.

[21] T. Hastie, R. Tibshirani, and J. Friedman. Unsupervised learning. In *The elements of statistical learning*, pages 485–585. Springer, 2009.

[22] S. Haykin. *Neural Networks and Learning Machines Third Edition*. 2008.

[23] J. Ho, D. W. Engels, and S. E. Sarma. Hiq: a hierarchical q-learning algorithm to solve the reader collision problem. In *International Symposium on Applications and the Internet Workshops (SAINTW'06)*, pages 4–pp. IEEE, 2006.

[24] T. Horiuchi, A. Fujino, O. Katai, and T. Sawaragi. Fuzzy interpolation-based q-learning with profit sharing plan scheme. In *Proceedings of 6th International Fuzzy Systems Conference*. Institute of Electrical & Electronics Engineers (IEEE), 1997.

[25] S. Kar, J. M. F. Moura, and H. V. Poor. Qd-learning: A collaborative distributed strategy for multi-agent reinforcement learning through consensus+ innovations. *IEEE Transactions on Signal Processing*, 61(7):1848–1862, apr 2013.

[26] H. M. La, R. Lim, and W. Sheng. Multirobot cooperative learning for predator avoidance. *IEEE Transactions on Control Systems Technology*, 23(1):52–63, jan 2015.

[27] E. A. Lee. The problem with threads. *Computer*, 39(5):33–42, 2006.

[28] J. Loch and S. P. Singh. Using eligibility traces to find the best memoryless policy in partially observable markov decision processes. In *ICML*, pages 323–331, 1998.

[29] L. MacDermed. Scaling up game theory: Achievable set methods for efficiently solving stochastic games of complete and incomplete information. In *Proceedings of the Twenty-Fifth AAAI Conference on Artificial Intelligence, AAAI 2011, San Francisco, California, USA, August 7-11, 2011*, 2011.

[30] T. Máhr, J. Srour, M. De Weerdt, and R. Zuidwijk. Can agents measure up? a comparative study of an agent-based and on-line optimization approach for a drayage problem with uncertainty. *Transportation Research Part C: Emerging Technologies*, 18(1):99–119, 2010.

[31] V. Mnih, K. Kavukcuoglu, D. Silver, A. Graves, I. Antonoglou, D. Wierstra, and M. A. Riedmiller. Playing atari with deep reinforcement learning. *CoRR*, abs/1312.5602, 2013.

[32] K. Morihiro, T. Isokawa, H. Nishimura, and N. Matsui. Characteristics of flocking behavior model by reinforcement learning scheme. In *2006 SICE-ICASE International Joint Conference*. Institute of Electrical & Electronics Engineers (IEEE), 2006.

[33] J. Oh. *Multiagent Social Learning in Large Repeated Games*. PhD thesis, Pittsburgh, PA, USA, 2009. AAI3414065.

[34] X. Pennec. Probabilities and statistics on riemannian manifolds: Basic tools for geometric measurements. In *NSIP*, pages 194–198. Citeseer, 1999.

[35] H. Prasad, P. LA, and S. Bhatnagar. Two-timescale algorithms for learning nash equilibria in general-sum stochastic games. In *Proceedings of the 2015 International Conference on Autonomous Agents and Multiagent Systems*, pages 1371–1379. International Foundation for Autonomous Agents and Multiagent Systems, 2015.

[36] C. W. Reynolds. Flocks, herds and schools: A distributed behavioral model. *ACM SIGGRAPH Computer Graphics*, 21(4):25–34, aug 1987.

[37] H. Saad, A. Mohamed, and T. ElBatt. Cooperative q-learning techniques for distributed online power allocation in femtocell networks. *Wirel. Commun. Mob. Comput.*, 15(15):1929–1944, feb 2014.

[38] J. Schulman, S. Levine, P. Abbeel, M. I. Jordan, and P. Moritz. Trust region policy optimization. In *Proceedings of the 32nd International Conference on Machine Learning, ICML 2015, Lille, France, 6-11 July 2015*, pages 1889–1897, 2015.

[39] Z. Shi, J. Tu, Q. Zhang, X. Zhang, and J. Wei. The improved q-learning algorithm based on pheromone mechanism for swarm robot system. In *Proceedings of the 32nd Chinese Control Conference*, pages 6033–6038, July 2013.

[40] D. Silver, A. Huang, C. J. Maddison, A. Guez, L. Sifre, G. Van Den Driessche, J. Schrittwieser, I. Antonoglou, V. Panneershelvam, M. Lanctot, et al. Mastering the game of go with deep neural networks and tree search. *Nature*, 529(7587):484–489, 2016.

[41] H. A. Simon. Why should machines learn? In *Machine learning*, pages 25–37. Springer, 1983.

[42] R. Sutton and A. Barto. Reinforcement learning: An introduction. *IEEE Trans. Neural Netw.*, 9(5):1054–1054, sep 1998.

[43] M. Tan. Multi-agent reinforcement learning: Independent versus cooperative agents. In *Machine Learning, Proceedings of the Tenth International Conference, University of Massachusetts, Amherst, MA, USA, June 27-29, 1993*, pages 330–337, 1993.

[44] B.-N. Wang, Y. Gao, Z.-Q. Chen, J.-Y. Xie, and S.-F. Chen. A two-layered multi-agent reinforcement learning model and algorithm. *Journal of Network and Computer Applications*, 30(4):1366–1376, nov 2007. Competitive.

[45] C. J. Watkins and P. Dayan. Q-learning. *Machine learning*, 8(3-4):279–292, 1992.

[46] P. Xuan, V. Lesser, and S. Zilberstein. Communication in multi-agent markov decision processes. In *MultiAgent Systems, 2000. Proceedings. Fourth International Conference on*, pages 467–468. IEEE, 2000.

[47] Z. Zhang, D. Zhao, J. Gao, D. Wang, and Y. Dai. FMRQ-a multiagent reinforcement learning algorithm for fully cooperative tasks. *IEEE Trans. Cybern.*, pages 1–13, 2016.

[48] M. Zolfpour-Arokhlo, A. Selamat, S. Z. M. Hashim, and H. Afkhami. Modeling of route planning system based on q value-based dynamic programming with multi-agent reinforcement learning algorithms. *Engineering Applications of Artificial Intelligence*, 29:163–177, mar 2014.